\documentclass{article}

\usepackage{arxiv}

\usepackage[utf8]{inputenc} 
\usepackage[T1]{fontenc}    
\usepackage{hyperref}       
\usepackage{url}            
\usepackage{booktabs}       
\usepackage{amsfonts}       
\usepackage{nicefrac}       
\usepackage{microtype}      
\usepackage{lipsum}		
\usepackage{graphicx,amssymb,amsmath,color}
\usepackage{doi}
\usepackage{array}
\newcolumntype{P}[1]{>{\centering\arraybackslash}p{#1}}

\usepackage[table,x11names]{xcolor}
\usepackage{dcolumn}
\usepackage{setspace}
\usepackage{rotating}
\usepackage{afterpage}
\usepackage[numbers,sort&compress]{natbib}

\newcolumntype{.}{D{.}{.}{-1}}

\usepackage{authblk}

\newcommand{\fig}{Fig.~}

\newcommand{\eqn}{Eq.~}
\newcommand{\eqns}{Eqs.~}

\newcommand{\sectn}{Section~}

\newcommand{\NS}{Navier--Stokes~}
\newcommand{\HS}{Hele--Shaw~}

\definecolor{blue}{rgb}{0, 0.5, 0.5}
\definecolor{blue2}{rgb}{0.1216, 0.4667, 0.7059}
\definecolor{red}{rgb}{0.8500, 0.1250, 0.0480} 
\definecolor{red2}{rgb}{0.8392, 0.1529, 0.1569} 
\definecolor{orange2}{rgb}{1.0, 0.498, 0.0549} 
\definecolor{yellow2}{rgb}{0.9290, 0.6940, 0.1250}
\definecolor{purple}{rgb}{0.4940, 0.1840, 0.5560}
\definecolor{purple2}{rgb}{0.5804, 0.4039, 0.7412}
\definecolor{green}{rgb}{0.4660, 0.6740, 0.1880}
\definecolor{green2}{rgb}{0.1725, 0.6275, 0.1725}
\definecolor{ltblue2}{rgb}{0.0902, 0.7451, 0.8118}
\definecolor{dkred2}{rgb}{0.6350, 0.0780, 0.1840}
\definecolor{gray2}{rgb}{0.22, 0.22, 0.3}
\definecolor{gray3}{rgb}{0.5, 0.5, 0.5}

\usepackage{mathtools}
\DeclarePairedDelimiter\bra{\langle}{\rvert}
\DeclarePairedDelimiter\ket{\lvert}{\rangle}
\DeclarePairedDelimiterX\braket[2]{\langle}{\rangle}{#1\,\delimsize\vert\,\mathopen{}#2}

\usepackage{tikz}
\usetikzlibrary{quantikz}

\title{Solving the Hele--Shaw flow using the Harrow--Hassidim--Lloyd algorithm on superconducting devices: A study of efficiency and challenges}

\author[a,$\dagger$]{Muralikrishnan Gopalakrishnan Meena}
\author[a]{Kalyana C. Gottiparthi}
\author[a,b]{Justin G. Lietz}
\author[a]{Antigoni Georgiadou}
\author[a]{Eduardo Antonio Coello P\'{e}rez}

\affil[a]{National Center for Computational Sciences, Oak Ridge National Laboratory, Oak Ridge, TN, USA}
\affil[b]{Current affiliation: NVIDIA, Quantum Architecture and Algorithm, Santa Clara, CA, USA}
\affil[$\dagger$]{To whom correspondence should be addressed: \href{mailto:gopalakrishm@ornl.gov}{gopalakrishm@ornl.gov}}

\date{}

\renewcommand{\headeright}{~}
\renewcommand{\undertitle}{~}
\renewcommand{\shorttitle}{Solving the Hele--Shaw flow using the HHL algorithm on superconducting devices}

\begin{document}

\maketitle

\begin{abstract}
The development of quantum processors for practical fluid flow problems is a promising yet distant goal. Recent advances in quantum linear solvers have highlighted their potential for classical fluid dynamics. In this study, we evaluate the Harrow--Hassidim--Lloyd (HHL) quantum linear systems algorithm (QLSA) for solving the idealized Hele--Shaw flow. Our focus is on the accuracy and computational cost of the HHL solver, which we find to be sensitive to the condition number, scaling exponentially with problem size. This emphasizes the need for preconditioning to enhance the practical use of QLSAs in fluid flow applications. Moreover, we perform shots-based simulations on quantum simulators and test the HHL solver on superconducting quantum devices, where noise, large circuit depths, and gate errors limit performance. Error suppression and mitigation techniques improve accuracy, suggesting that such fluid flow problems can benchmark noise mitigation efforts. Our findings provide a foundation for future, more complex application of QLSAs in fluid flow simulations.

\textbf{Notice:} This manuscript has been authored by UT-Battelle, LLC, under contract DE-AC05-00OR22725 with the US Department of Energy (DOE). The US government retains and the publisher, by accepting the article for publication, acknowledges that the US government retains a nonexclusive, paid-up, irrevocable, worldwide license to publish or reproduce the published form of this manuscript, or allow others to do so, for US government purposes. DOE will provide public access to these results of federally sponsored research in accordance with the DOE Public Access Plan (\href{http://energy.gov/downloads/doe-public-access-plan}{http://energy.gov/downloads/doe-public-access-plan}).
\end{abstract}


\section{Introduction}
\label{sec:intro}

Fluid dynamics is fundamental to almost all phenomena in nature and engineering where fluids are present. These include vast applications such as complex systems in Earth science (e.g. atmosphere and ocean turbulence), engineering (e.g. gas turbines, wind turbines, aerodynamics), and astrophysics (e.g. stellar evolution, supernovae, planetary atmosphere). What makes analyzing and modeling fluid flows challenging is their nonlinear behavior. The dynamics of fluid flows is governed by a set of multi-variate, multi-physics, higher-order, nonlinear partial differential equations (PDEs) - the Navier--Stokes equations. Numerically solving these PDEs for real-world applications can be extremely computationally expensive, even with the latest high-performance computing systems \cite{norman2021unprecedented}. A wide range of computational techniques exist to handle nonlinear PDEs using classical computing resources. Iterative solvers using implicit or explicit time stepping have been a popular choice with spatial discretization done using finite difference or finite volume. These techniques for fluid dynamics, turbulence, and combustion are available in literature, and an interested reader can find these methods summarized here \cite{Pitsch_2006, Mani_2023}.

The advent of various linear \cite{harrow2009quantum,childs2021high} and nonlinear \cite{leyton2008quantum,lubasch2020variational} PDE solver techniques in quantum computing \cite{nielsen2010quantum} provides a great avenue to tackle fluid flow problems with potentially exponential speed-ups. Note that while classical computing techniques use algorithms which are of order $\mathcal{O}({n^m\log(n)})$ or $\mathcal{O}({n^m})$, where $m$ is a positive integer and $n$ the problem size, quantum computing has the potential for $\mathcal{O}(\log{n})$. The first step towards such usage is to efficiently solve fundamental fluid flow models, enabling concrete foundation for modeling complex behavior of fluid flows and showcase promising potential for quantum methods to solve hard problems.

While tackling the nonlinear PDE is of utmost importance to answer many unsolved fluid flow problems, the \NS equations in their reduced complexity form -- with various approximations of time evolution and nonlinearities -- are important to building fundamental models. These flows with reduced complexity are generally termed as ideal fluid flows. Generally, the governing equations to these flows are linearized versions of the \NS, and thus, the solutions can usually be obtained analytically, if not numerically using much simpler techniques compared to solving the full \NS equations.

\begin{figure}
\begin{center}
  \includegraphics[width=0.35\textwidth]{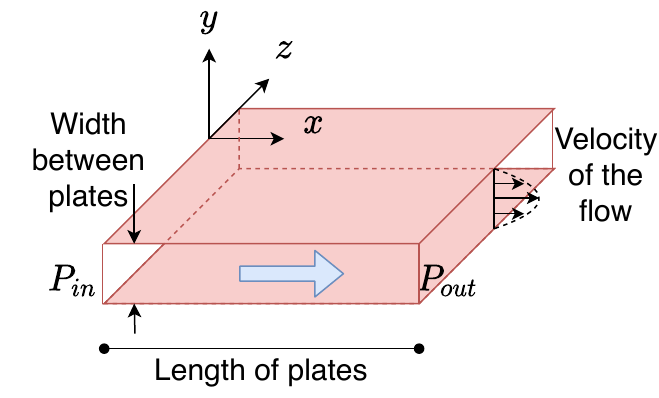}
\end{center}
 \caption{Sample portrayal of the Hele-Shaw flow representing the flow between two flat plates driven by pressure difference at inlet and outlet.
\label{fig:H-S_flow}}
\end{figure}

Although most ideal fluid flows are analytical models, some have direct applications in engineering. One such flow is the Hele--Shaw flow problem \cite{hele1898flow}. The \HS flow comprises of flow between two flat plates separated by infinitesimally small distance (see \fig\ref{fig:H-S_flow}), enabling approximations in the \NS equations. The solution to the resulting linear system of equations governing the velocity and pressure profiles can be obtained analytically, which are parabolic and linear profiles, respectively. We note that the \HS flow is not chosen for this study because it is a challenging fluid flow problem for classical computation, but rather because it can be analyzed using currently available quantum computing resources, serving as a sample benchmark problem. Moreover, the flow also has very important implications to real-world applications involving micro-flows where such geometric scenarios are prevalent (e.g. microfluidics, groundwater flow, modeling geological processes, porous media flow, oil recovery, and bioengineering). 

In the past years there have been many demonstrations of solving linear system of equations governing fluid flows and linearized versions of the \NS using quantum algorithms \cite{yepez2001quantum,cao2013quantum,todorova2020quantum,budinski2021quantum,oz2022solving,bharadwaj2023hybrid,oz2023efficient}. Recently a solution to the \HS flow problem using a quantum linear solver algorithm (QLSA) was demonstrated \cite{bharadwaj2020quantum}. Some of these demonstrations have been performed both on quantum simulators and on quantum hardware, such as IBM's superconducting quantum devices.

While these demonstrations show the analytical and numerical promise of quantum computing techniques, limited work has been done in assessing the efficiency and practical challenges involved in solving these PDEs. Such assessments are needed while pathfinding quantum implementations for problems of practical utility \cite{herrmann2023quantum}. In this work, we focus on a preliminary evaluation of the efficiency of solving the \HS problem using a canonical QLSA, the Harrow--Hassidim--Lloyd (HHL) algorithm \cite{harrow2009quantum}. Furthermore, we describe various challenges in solving the problem, specifically on superconducting quantum devices of IBM. We note that there are many other QLSA algorithms \cite{childs2017quantum,low2017optimal,gilyen2019quantum,subacsi2019quantum,an2020quantum,lin2022heisenberg,tong2021fast,dong2021random}, including variational (hybrid quantum-classical) methods \cite{bravo2023variational}, which have shown better performance than the HHL algorithm on current noisy intermediate-scale quantum (NISQ) devices. The objective of this study is to provide a preliminary assessment of a canonical QLSA for solving an ideal fluid flow problem and offer guidance for evaluating state-of-the-art QLSAs. 

The primary contributions of this study are as follows:
\begin{enumerate}
    \item We present a preliminary assessment of the efficiency and practical challenges involved in solving the \HS flow (an idealized fluid flow problem) using the HHL algorithm (a canonical QLSA) on superconducting devices. We hope that these assessments can serve as a guidance on evaluating more complex fluid flow problems using various state-of-the-art QLSAs.
    \item We report the computational scalability and accuracy metrics with respect to change in problem size.
    \item We demonstrate noise modeling using simulators and comparison of results with that from devices.
    \item We report the effect of various error suppression and mitigation techniques on the accuracy of results from devices.
\end{enumerate}

In what follows, we first describe the governing equations of the \HS flow problem, the analytical solution, and the QLSA in the Methodology section, \sectn\ref{sec:theory}. The results and observations of our analyses are presented in \sectn\ref{sec:results}. Some concluding remarks and future directions are discussed in \sectn\ref{sec:conclusion}.

\section{Methodology}
\label{sec:theory}

\begin{figure*}
\begin{center}
  \includegraphics[width=0.8\textwidth]{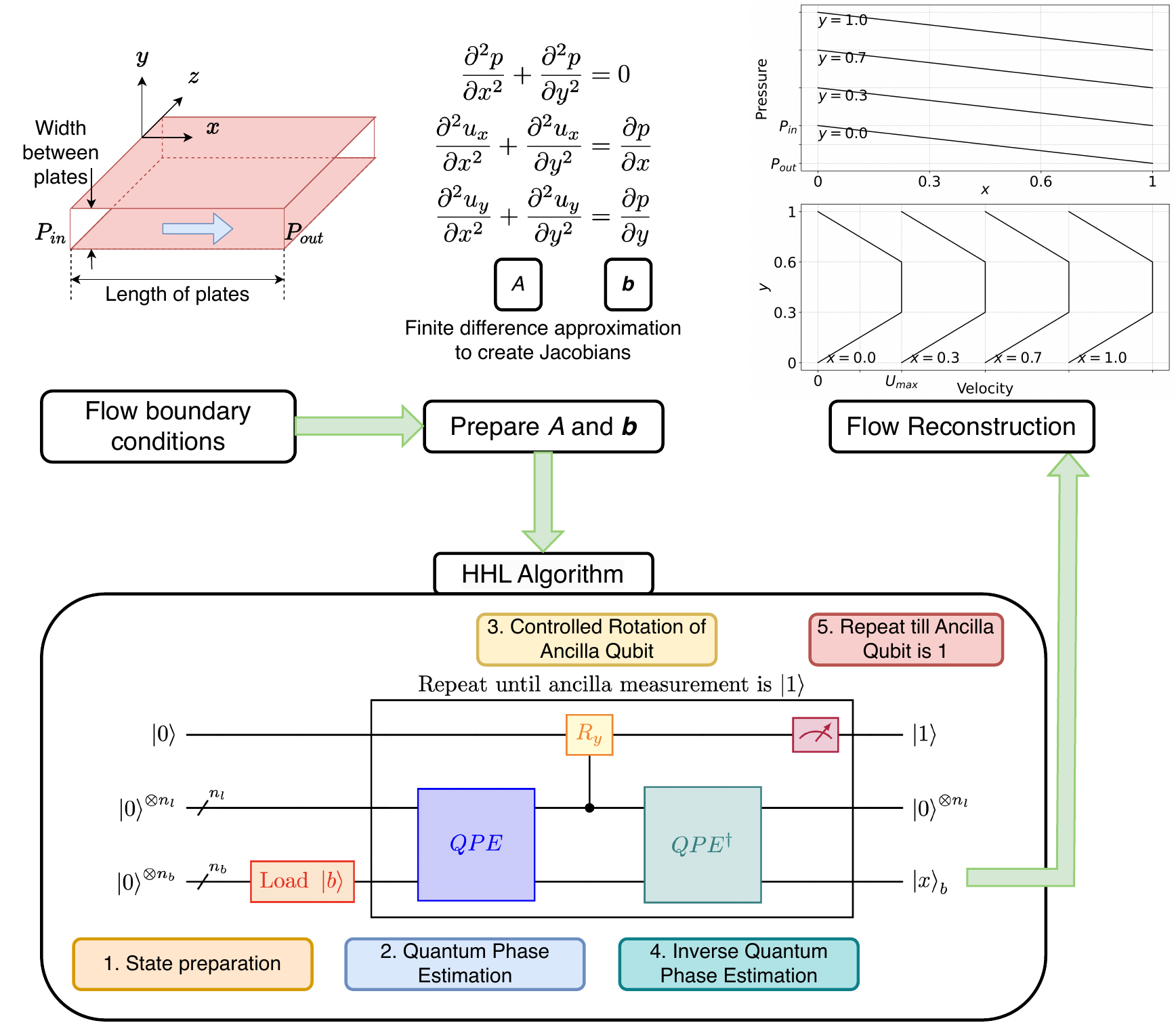}
\end{center}
\caption{Outline of the methodology for solving the \HS flow using the HHL algorithm.
\label{fig:outline}}
\end{figure*}

An overview of the methodology is shown in \fig\ref{fig:outline}. Given the initial and/or boundary conditions of the flow, we formulate a linear system of equations in the form $A\boldsymbol{x}=\boldsymbol{b}$ using the governing equations of the fluid flow problem. A finite difference approximation is employed to construct the $A$ matrix and $\boldsymbol{b}$ vector, which are then provided as inputs to the HHL algorithm. The output of the HHL algorithm is the state vector $\boldsymbol{x}$, representing the flow variables, which is used to reconstruct the flow field. Below, we first describe the flow problem addressed in this study, the \HS flow, followed by a detailed explanation of the HHL algorithm.

\subsection{Hele--Shaw flow setup}

The \HS flow is steady (a time independent flow) and two-dimensional. The governing equations are obtained by imposing incompressibility and inviscid approximations to the Navier--Stokes equations, resulting in
\begin{align}
    \frac{\partial u_x}{\partial x} + \frac{\partial u_y}{\partial y} &= 0\label{eq:H-S_m}\\
    \frac{\partial^2 u_x}{\partial x^2} + \frac{\partial^2 u_x}{\partial y^2} - \frac{\partial p}{\partial x} &= 0\label{eq:H-S_ux}\\
    \frac{\partial^2 u_y}{\partial x^2} + \frac{\partial^2 u_y}{\partial y^2} - \frac{\partial p}{\partial y} &= 0\label{eq:H-S_uy}
\end{align}
where $x$ and $y$ are the horizontal and vertical directions, respectively. $u_x$ and $u_y$ represent the velocity of the flow in the horizontal and vertical directions, and $p$ represents the pressure in the flow field. The length, velocity, time, and pressure variables are non-dimensionalized by the length of the flat plates, $L$, the maximum velocity of the flow, $U$, which is at the centerline between the flat plates, the ratio of $L/U$, and $\mu U/L$, respectively. Here $\mu$ represents the viscosity of the fluid. The Reynolds number of this idealized flow, measured using the width of the plates, $D$, is assumed to approach zero as the width becomes infinitesimally small. We choose $\mu=1$, $L=1$, pressure at the inlet $P_{in}=200$, pressure at the outlet $P_{out}=0$, and $D=L$.

To obtain the governing equations in the form of a linear system of equations, $A\boldsymbol{x}=\boldsymbol{b}$, the velocity and pressure terms are decoupled by taking the divergence of \eqns\ref{eq:H-S_ux} and \ref{eq:H-S_uy}, and applying \eqn\ref{eq:H-S_m}, resulting in
\begin{align}
    \frac{\partial^2p}{\partial x^2} + \frac{\partial^2p}{\partial y^2} &= 0\label{eq:H-S_linear_p}\\
    \frac{\partial^2 u_x}{\partial x^2} + \frac{\partial^2 u_x}{\partial y^2} &= \frac{\partial p}{\partial x} \label{eq:H-S_linear_ux}\\
    \frac{\partial^2 u_y}{\partial x^2} + \frac{\partial^2 u_y}{\partial y^2} &= \frac{\partial p}{\partial y}\label{eq:H-S_linear_uy}.
\end{align}

The above Poisson equations can be solved analytically to obtain the solutions
\begin{align}
    p &= \frac{\left(P_{in} - P_{out}\right)}{L}x + P_{in}\label{eq:H-S_analytical_p}\\
    u_x &= \frac{1}{2\mu}\frac{\left(P_{out} - P_{in}\right)}{L}y\left(D-y\right)\label{eq:H-S_analytical_ux}\\
    u_y &= 0\nonumber.
\end{align}
Since the vertical velocity, $u_y$, is zero everywhere, we focus on solving the coupled system of linear equations for pressure and horizontal velocity, \eqns\ref{eq:H-S_linear_p} and \ref{eq:H-S_linear_ux}. 

Classically, numerical solutions to these equations can be obtained by discretizing them on a grid using finite difference approximations and solving the resulting Jacobians in an iterative process -- first solving for pressure and using the solution to solve for velocity. We use various numbers of grid points in the $x$ and $y$ directions to assess the efficiency of the QLSA. To obtain the Jacobian matrices for $p$ and $u_x$, we used second-order finite difference approximation (first order at the boundaries) to discretize the flow field. For the quantum solution, the two equations are solved separately using the QLSA. To solve for $u_x$, the efficiency of the QLSA was tested using both analytical and quantum solution of $p$, the latter addressing the coupled system.

\subsection{Quantum linear solver algorithm}

\begin{figure*}
\begin{center}
  \includegraphics[width=0.8\textwidth]{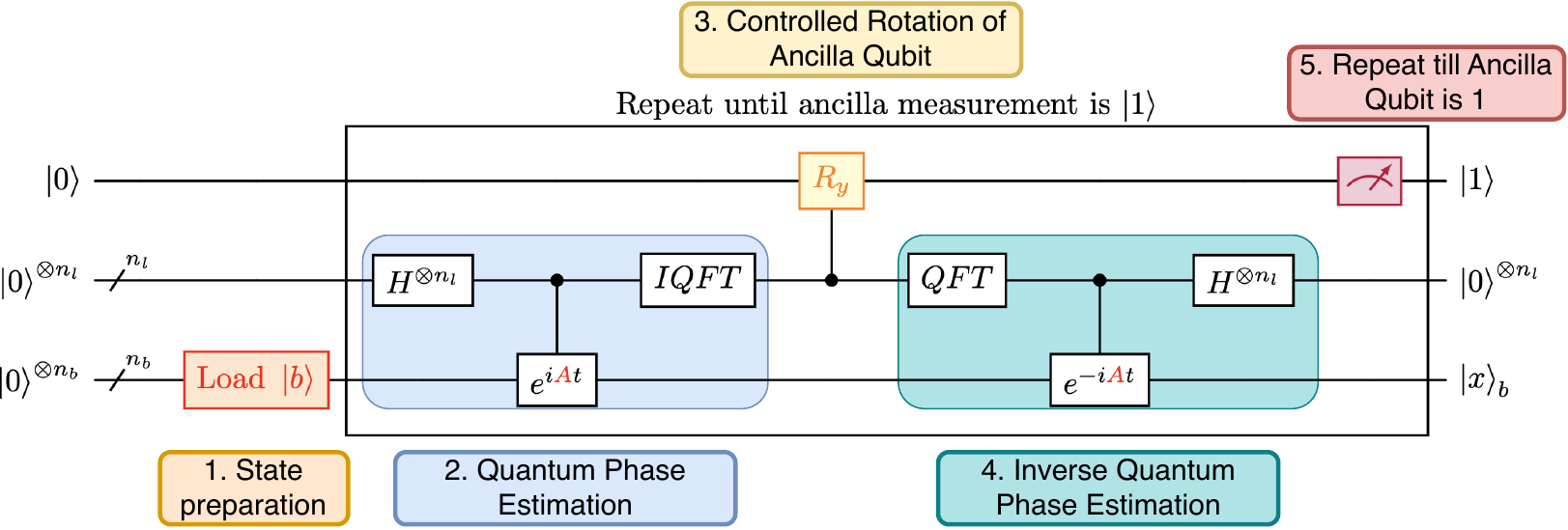}
\end{center}
\caption{A sample quantum circuit depicting the HHL algorithm \cite{harrow2009quantum}. The algorithm is implemented using Qiskit.
\label{fig:HHL}}
\end{figure*}

We use the Harrow--Hassidim--Lloyd (HHL) \cite{harrow2009quantum} algorithm to solve \eqns\ref{eq:H-S_linear_p} and \ref{eq:H-S_linear_ux}. The HHL algorithm is one of the first QLSAs to be introduced. The HHL algorithm fundamentally relies on quantum phase estimation (QPE) \cite{nielsen2010quantum,lin2022lecture} to obtain the solutions. Various other QLSA implementations using different quantum algorithms have been introduced with the aim of improving computational performance and accuracy. These include the use of linear combination of unitaries \cite{childs2017quantum}, quantum signal processing \cite{low2017optimal}, quantum singular value transformation \cite{gilyen2019quantum}, randomized method based on adiabatic computing \cite{subacsi2019quantum}, near-optimal eigenstate filtering \cite{lin2020near}, and fast inversion (preconditioning) \cite{tong2021fast}. In this preliminary work, we focus our efforts on assessing the capability of QLSA for solving ideal fluid flows using the HHL algorithm. A sample portrayal of the HHL algorithm is provided in \fig\ref{fig:HHL}.


Given a system of linear equations defined by states $\boldsymbol{x},\boldsymbol{b}\in\mathbb{C}^N$ and a Hermitian and invertible system matrix $A\in\mathbb{C}^{N\times N}$ (usually the Jacobian in flow problems), the overall methodology is to prepare a quantum state $\ket{\boldsymbol{x}}$ representing the solution $\boldsymbol{x}$ in terms of the eigenbasis of the system matrix $A$. The HHL algorithm mainly has five steps: state preparation, quantum phase estimation (QPE), ancilla bit rotation, inverse quantum phase estimation (IQPE), and measurement.

The first step is to encode the available information into the quantum states (state preparation). This can be achieved, for example, via normalization of states and the amplitude-embedding technique~\cite{weigold2020data}. In terms of the quantum states, the problem reads
\begin{equation}
    \ket{\boldsymbol{x}}   = A^{-1} \ket{\boldsymbol{b}}.
\end{equation}
Using the spectral decomposition of $A$ and representing $\ket{\boldsymbol{b}}$ in terms of the resulting eigenbasis, we obtain
\begin{align}
    A          &= \sum_{j}^{N} \lambda_j \ket{u_j} \bra{u_j}\\
    \ket{\boldsymbol{b}}    &= \sum_{j}^{N}b_j \ket{u_j}
\end{align}
where $\lambda_j\in\mathbb{R}$ and $\ket{u_j}$ are the $j$th eigenvalue and eigenvector of $A$, respectively, and $b_j\in\mathbb{C}$. 
Thus, the solution quantum state can be written as
\begin{equation}
    \ket{\boldsymbol{x}} = \sum_{j}^{N}\lambda_j^{-1}b_j\ket{u_j}.
\end{equation}

After preparing the quantum state $\ket{\boldsymbol{b}}$ using $n_b=\log_2N$ qubits, eigenvalue estimation is performed using QPE. QPE has three components: superposition of the $n_l$ qubits using Hadamard gates $H$, controlled rotation using a unitary gate representing the $A$ matrix, and inverse quantum Fourier transform (IQFT). The $A$ matrix is encoded as the Hamiltonian of a unitary gate $U=e^{iAt}$. The overall objective of QPE is to extract the phase of the eigenvalues of $U$ and store them in the $n_l$ qubits. This is achieved after the IQFT step. The accuracy of the estimation depends on how accurately the phase can be stored, dependent on the value of $n_l$. For a detailed description and walk through of the algorithm, we recommend the reader to follow \cite{dervovic2018quantum,zaman2023step}.

Next, an ancilla qubit is rotated based on the eigenvalues stored in the $n_l$ qubits using the controlled rotation gate $R_y$. At this stage, if the ancilla qubit is measured, it will collapse to either $\ket{0}$ or $\ket{1}$. If the measurement is $\ket{0}$, the results are discarded and the circuit is repeated until $\ket{1}$ is measured and the solution vector will be stored in the $n_b$ qubit registers. However, we cannot obtain the solution vector in terms of $\ket{0}/\ket{1}$ since the $n_b$ registers are entangled with the $n_l$ registers. Thus, the state is uncomputed through the IQPE step, $QPE^\dagger$, during which the $n_b$ and $n_l$ qubits are unentangled and the solution vector $\ket{x}$ is stored in the $n_b$ registers as $\ket{0}/\ket{1}$ measurements. This step also results in the $n_l$ qubits being reset to $\ket{0}$. Thus the final desired measurement is $\ket{1}$ for the ancilla qubit, $\ket{0}$ for the $n_l$ registers, and $\ket{0}/\ket{1}$ for the $n_b$ registers.

Note that the ancilla qubit is measured to obtain a function approximation of the solution vector, like the Euclidean norm, $L_2$, of the vector. We can also measure the actual components of the vector $\ket{\boldsymbol{x}}$ -- when the outcome of the ancilla qubit is $\ket{1}$, the state registers correspond to the full solution vector. Extracting the Euclidean norm is more computationally efficient and might be of interest to certain fluid flow problems where global metrics such as total energy, dissipation rate, and flow mixing efficiency are of interest. Most fluid flow problems require extracting the solution state, although, the process can hamper the computational speedup attained from the QLSA algorithm.

We use Qiskit \cite{Qiskit} (version $0.44.1$), an open-source SDK for quantum computing, to perform the analysis. For the Qiskit implementation used in this work\footnote{\href{https://github.com/anedumla/quantum_linear_solvers}{https://github.com/anedumla/quantum\_linear\_solvers}}, a total of $n_b + n_l + 1 + 2 = 2n_b + 3$ qubits are used to generate the HHL circuit, with the last two qubits used for representing sign of the eigenvalues and accuracy tolerance. State preparation is performed using isometry \cite{iten2016quantum} for the vector and Trotterization \cite{hatano2005finding} for the matrix.

The HHL algorithm estimates a function of the solution vector $\boldsymbol{x}$ with a computational complexity of  $\mathcal{O}(\log(N)s^{2}\kappa^{2}/\epsilon)$. Here, $s$ is the sparsity of $A$, $\kappa$ is the condition number of the system, and $\epsilon$ the accuracy of the approximation. The assumption is that we have efficient oracles for encoding the data, Hamiltonian simulation, and computing a function of the solution. Note that for preserving the exponential speed up of the HHL algorithm only a function approximation of the solution vector can be found by the algorithm. Measuring the actual state of $\ket{\boldsymbol{x}}$ would require time in linear complexity, leading to reduction in the speed up obtained by the algorithm.

An issue with the HHL algorithm for the fluid flow problem is that $A$ is rarely Hermitian. Thus, the complex conjugate of the matrix is used to generate a Hermitian matrix, and $\boldsymbol{b}$ is padded with zeros. This operation leads to increase in the system size from $N$ to $2N$, requiring an additional two qubits to generate the HHL circuit. 

Another bottleneck is that if the system size is not a power of $2$, i.e. when the number of grid points in the flow field is not a power of $2$, we need to pad $\boldsymbol{b}$ and $A$ with zeros. Additionally, ones are added on the corresponding diagonal elements of $A$ so that the matrix is not singular.


\section{Results and Discussion}
\label{sec:results}

We assess the efficiency and challenges of running the HHL circuit for the \HS problem on simulators and real quantum hardware -- superconducting devices of IBM.

\paragraph{Simulator setup} We use Qiskit's Aer simulator (version $0.13.3$) to validate the HHL implementation for the \HS problem. We use the statevector backend to run the simulation. Qiskit provides the capability to extract the states when using the statevector simulator, rather than just measuring the observable like the $L_2$ norm of the state. We also perform shot-based runs and obtain the quasi probabilities of measuring each bitstring. The states are recovered by re-normalizing these measurements.

\paragraph{Real hardware} We use the IBM Quantum Computing platform to run the circuits on real hardware. After transpiling the circuits, Qiskit's {\tt qiskit-ibm-runtime} (version $0.17.0$) is used to select IBM's superconducting devices as the real backend to run the circuits. Specifically, we use the IBM Nairobi and Cairo machines to run our circuits.

We take a systematic, incremental approach by first assessing the capability of simulators to run the circuit in the absence and presence of artificial noise, before running on real hardware. We explore the following tasks: (a) accuracy of the algorithm run on simulators, (b) computational cost of running simulators, (c) modeling and mitigating noise using simulators, and (d) results of running on real hardware.

\subsection{Accuracy of the algorithm}

\begin{figure*}
\begin{center}
\includegraphics[width=1\textwidth]{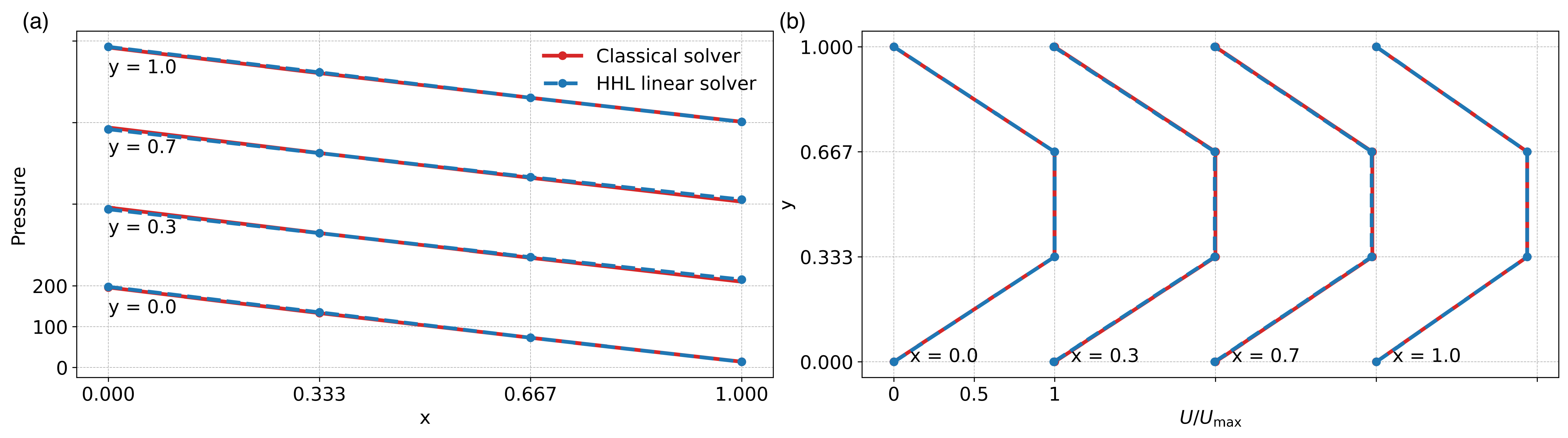}
\end{center}
\caption{(a) Pressure and (b) velocity profiles of the flow at various spatial locations in the domain. Pressure profiles across horizontal length are plotted at various vertical locations. Velocity profiles across vertical heights are plotted at various horizontal locations. Solution from a classical solver and the state reconstruction from the solution of the HHL algorithm are shown.
\label{fig:results_pressure-velocity}}
\end{figure*}

We first test the accuracy of the HHL algorithm to solve the \HS problem using simulators. We choose a domain size of $4 \times 4$, totaling $16$ grid points in the domain, including the boundary points. To solve this problem, the HHL circuit requires $11$ qubits ($n_b=4$, $n_l=4$, $1$ ancilla, and $2$ additional). The results of using the HHL algorithm to solve for the $p$ and $u_x$ profiles are presented in \fig\ref{fig:results_pressure-velocity}. The implementation achieves a state fidelity \cite{nielsen2010quantum} of $99.99\%$ for $p$ and $99.9995\%$ for $u_x$ when using the analytical pressure profile. The state fidelity of $u_x$ is $99.993\%$ when using the $p$ profile obtained using the HHL solver. We note that these results are comparable to existing implementations in literature \cite{bharadwaj2020quantum} and the latest Qiskit implementation of HHL achieves about $10\%$ higher fidelity than before.

\begin{figure}
\begin{center}
  \includegraphics[width=0.5\textwidth]{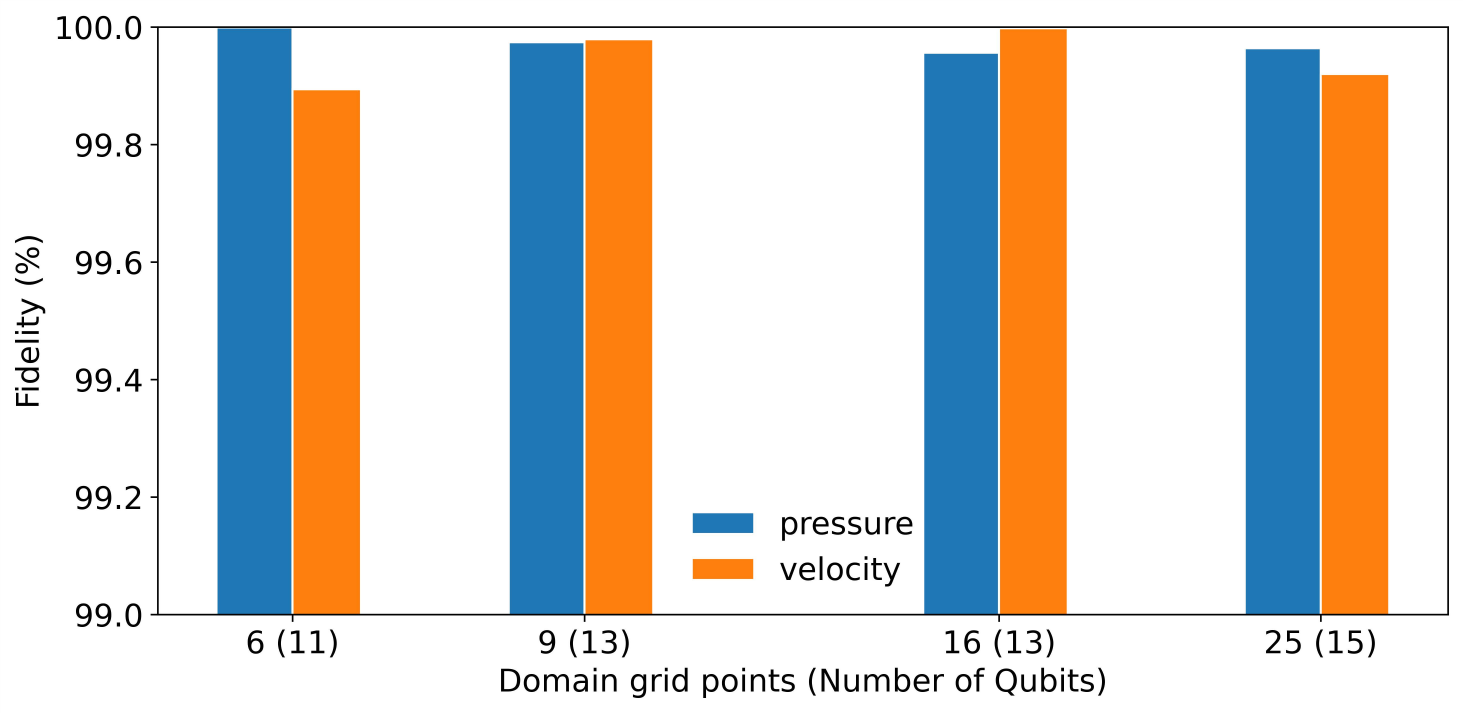}
\end{center}
 \caption{Scaling the fidelity of the HHL solver with varying domain grid points. 
\label{fig:results_pressure-velocity_fidelity-scale}}
\end{figure}

To test the scalability of the HHL algorithm, we vary the domain size and solve the corresponding system of linear equations. The state fidelity of the results are shown in \fig\ref{fig:results_pressure-velocity_fidelity-scale}. The general trend of the accuracy remains constant with less than $0.1\%$ and $0.05\%$ variability for velocity and pressure, respectively.

\subsection{Computational cost of simulators}

\begin{figure}
\begin{center}
  \includegraphics[width=0.5\textwidth]{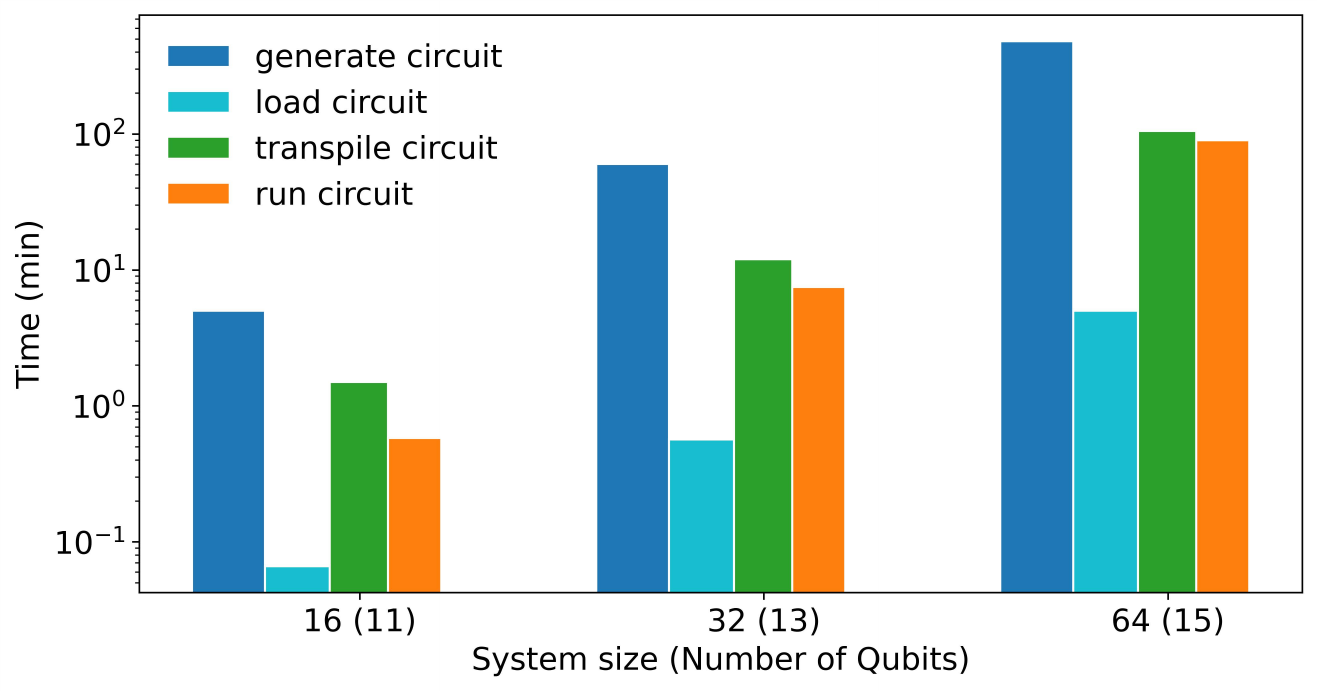}
\end{center}
 \caption{Scaling the simulator with varying system sizes. Different aspects of the solver workflow are timed.
\label{fig:results_simulator-scaling}}
\end{figure}

Next, we assess the computational capability of Qiskit's Aer simulator to solve the HHL circuits for the \HS problem. We measure the time taken at various steps in the solver workflow: generating the circuit using basis gates, transpilation of the circuit to match the topology of a given backend, and running the transpiled circuit using the backend. The results for varying domain size are shown in \fig\ref{fig:results_simulator-scaling}. The total execution time of the simulators is quite high. A significant portion of the time is consumed by generating the HHL circuit for the given size of the \HS problem. For this reason, we found that saving and loading the circuits to be helpful for studies independent of the circuit generation process, such as for testing various backends, debugging measurement procedure, and noise modeling and mitigation strategies.

We note that the times for all the parts of the workflow scales exponentially with the domain size. These results suggest the need for more effective procedures to generate the HHL circuit and potential use of GPU-based backends to accelerate the simulations. Qiskit does provide a multi-node GPU-based Aer simulator, using cuQuantum \cite{bayraktar2023cuquantum}. However, a key point to note here is that the problem in question does not strong-scale and the current problem sizes do not fill up the GPU memory to take advantage of the parallel implementation. The limiting factor is the process of generating the circuit.

\begin{figure}
\begin{center}
  \includegraphics[width=0.5\textwidth]{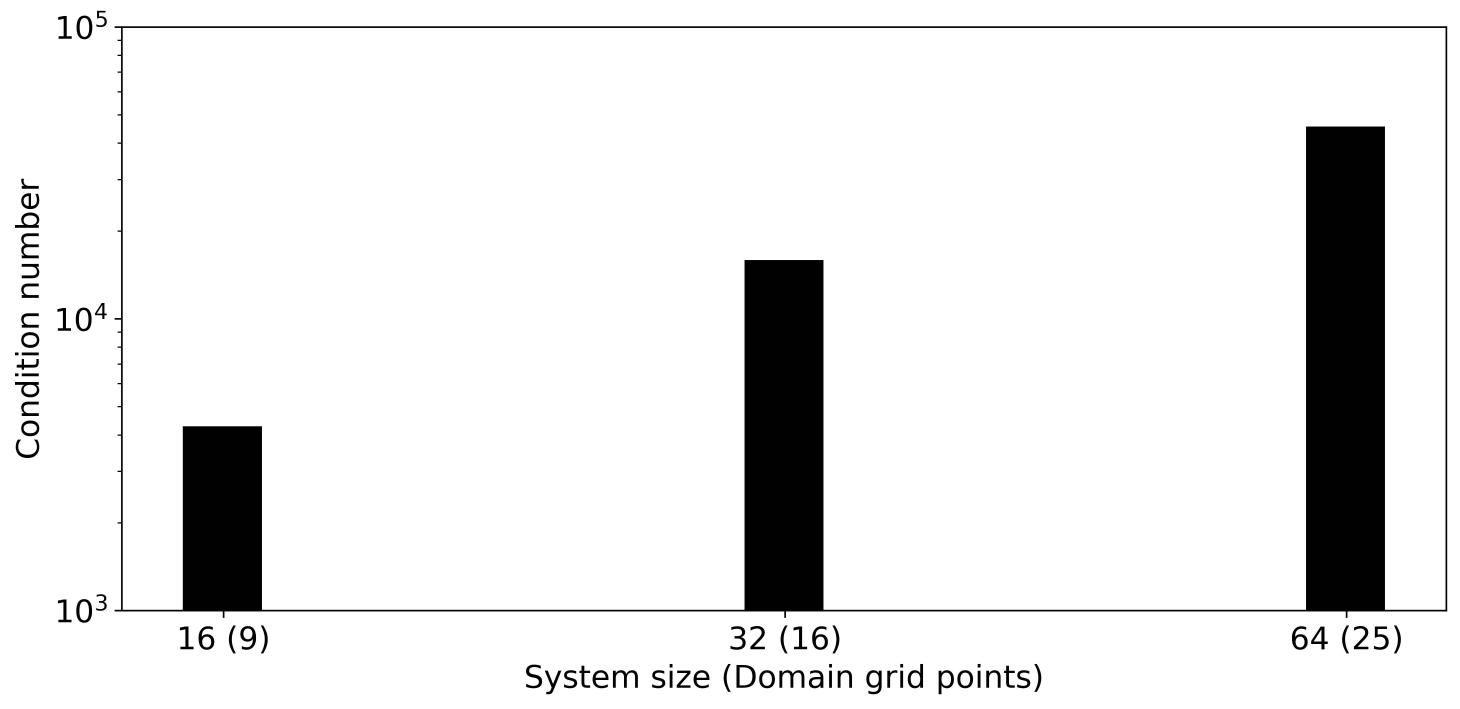}
\end{center}
 \caption{Scaling of the condition number of the problem with varying system sizes.
\label{fig:results_simulator-conditionnum}}
\end{figure}

It is important to note that all QLSAs have at least an $\mathcal{O}(\kappa)$ scaling, where $\kappa$ is the condition number of the system. As noted before, the HHL algorithm has an $\mathcal{O}(\kappa^2)$ scaling. Thus, it is important to analyze the $\kappa$ of the \HS problem at various grid resolutions. The change in $\kappa$ for the problem with system size is shown in \fig\ref{fig:results_simulator-conditionnum}. We note that the $\kappa$ of the system scales exponentially with system size, which contributes to the high computational expense of the solving the problem using the HHL algorithm. Thus, preconditioning the problem is potentially a need for practical utility of the QLSA for solving such problems \cite{clader2013preconditioned,golden2022quantum}.

\subsection{Noise modeling and mitigation using simulators}

The current state-of-the-art quantum devices being noisy intermediate-scale quantum (NISQ) computers demand procedures to address the noise in the hardware. There are many possible errors which can be incurred during quantum computing, including those due to imperfect implementation of the gate operations, decoherence of qubits to outside environment, and measurement errors. To alleviate these errors, error suppression and mitigation strategies are generally used for NISQ devices. Performing these strategies on devices can be expensive. Thus, prior to running the circuits on real hardware, it is helpful to test the circuits on noisy simulators that model the effects of noise in the hardware. This helps assess the capabilities of noise suppression and mitigation strategies to alleviate the errors resulting from the hardware. For these purposes, we use Qiskit's \textit{fake} backends which model the noise of IBM devices from the average errors associated to the operation of the basic gates on each qubit, those gates' execution times, and the qubits' average relaxation and dephasing times ($T_1$ and $T_2$), together with average measurement errors. The Aer Simulator assumes that every error is uncorrelated to previous ones and that they can be represented as a simple bit-flip or phase-flip circuit instruction. The simulator then creates new quantum circuits for every instance an error can take place and associates occurrence probabilities to them. The quasi probabilities of measuring each bitstring are then sampled from this weighted list of quantum circuits.

\begin{figure*}
\begin{center}
\includegraphics[width=1\textwidth]{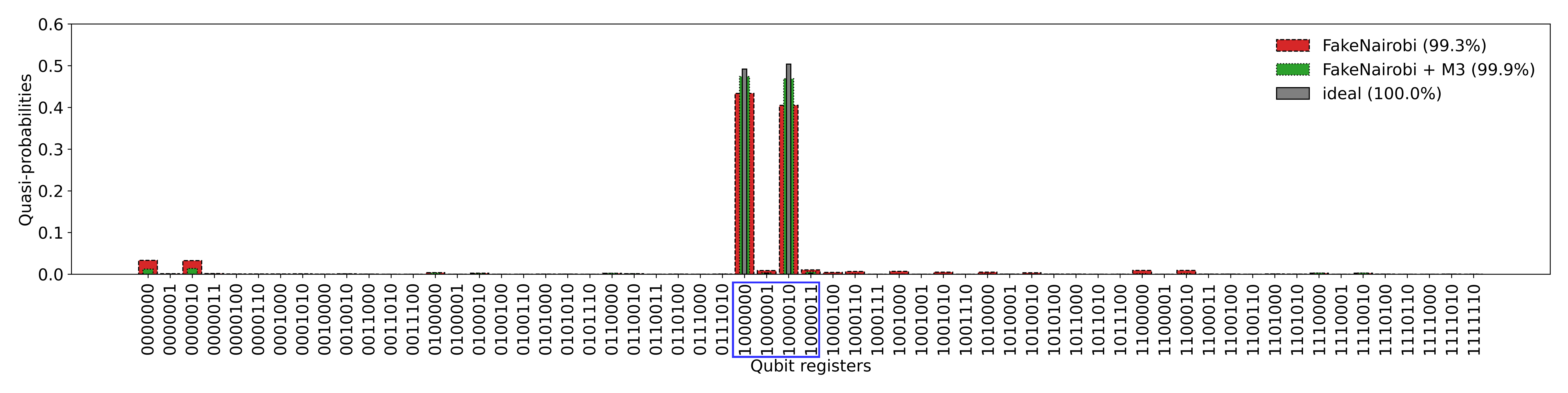}
\end{center}
 \caption{Noise modeling and mitigation study for the QLSA, demonstrated through solving the \HS problem with 4 grid points (size of matrix $A = [4\times4]$). Results of the shot count measurements (normalized to quasi-probabilities) for a run with 10,000 shots using (1) an ideal simulator -- Qiskit's Aer simulator with the {\tt statevector} backend, (2) a fake backend having noise model of the IBM Nairobi machine, and (3) the fake backend with qubit measurement error mitigation using the M3 technique. The measurements highlighted in the $x$-axis using the box (1000000, 1000001, 1000010, 1000011) are the ones corresponding to the solution vector. The fidelity of the results from the three simulations are also noted in the legend.
\label{fig:results_error_HS_ideal-fake}}
\end{figure*}

The results of running the HHL circuits for the \HS problem on Qiskit's {\tt FakeNairobi} backend are shown in \fig\ref{fig:results_error_HS_ideal-fake}, with the register measurements representing the solution state highlighted by the box. To compare the results, since the real backends are not capable of measuring the state of the system, we perform shot-based measurements on the fake backends to measure the quasi-probabilities of the qubit states and extract the registers that correspond to the solution state -- when ancilla qubit is 1 (the ones highlighted by the box in \fig\ref{fig:results_error_HS_ideal-fake}). The noisy simulator introduces various errors in the measurements, which is expected to model the behavior of the real backend.

We now try an error mitigation strategy to alleviate the noise introduced by the fake backend. We mainly try to address the errors related to the measurement of qubits. We use the matrix-free measurement mitigation (M3) \cite{nation2021scalable} technique implemented by Qiskit to test qubit measurement error mitigation. Results of applying the M3 technique on the fake backend simulation are shown in \fig\ref{fig:results_error_HS_ideal-fake} -- green bars. Both the individual quasi-probabilities and the state fidelity improve with the application of M3 technique. We are now positioned to test the capability of real hardware to run our circuits. We try additional error suppression strategies based on our experience with running the circuits on real hardware -- discussed in the next section.

\subsection{Running on real hardware}

\begin{figure*}
\begin{center}  \includegraphics[width=1\textwidth]{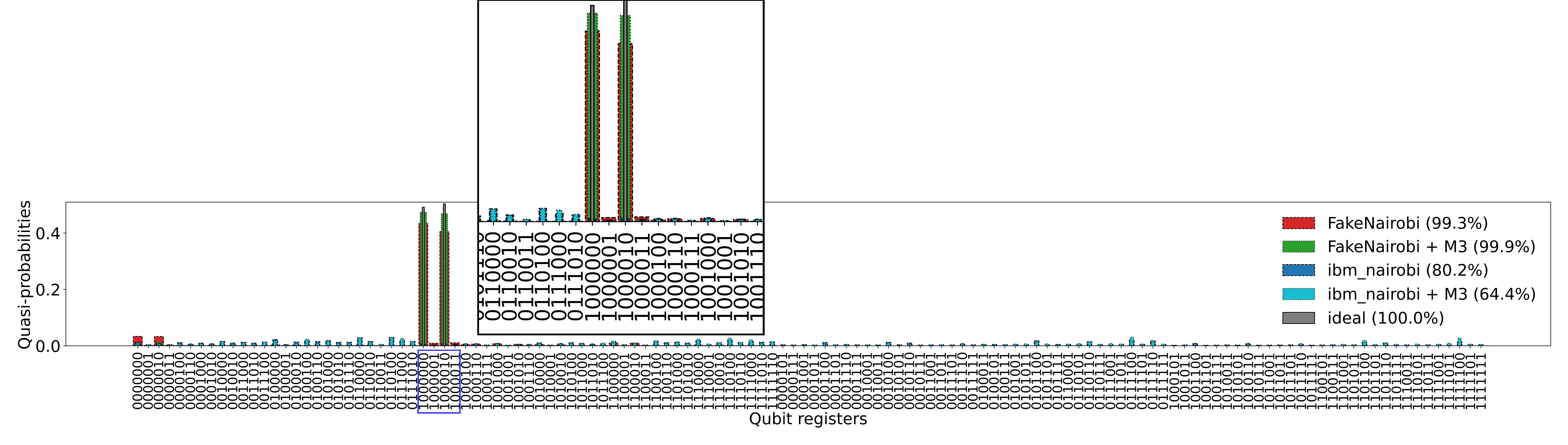}
\end{center}
 \caption{Results of running the \HS fluid flow problem analyzed in \fig\ref{fig:results_error_HS_ideal-fake} on a real backend. Results of the quasi-probabilities for a run with 10,000 shots using (1) an ideal simulator -- Qiskit's Aer simulator with the {\tt statevector} backend; (2) a fake backend having noise model of the IBM Nairobi machine, (3) the fake backend with M3 qubit measurement error mitigation, (4) real IBM Nairobi device, and (5) real IBM Nairobi device with M3 mitigation. The measurements highlighted in the $x$-axis using the box (1000000, 1000001, 1000010, 1000011) are the ones corresponding to the solution vector. A zoomed-in view of the relevant measurements are shown in the in-set plot. The fidelity of the results from the three simulations are also noted in the legend.
\label{fig:results_error_HS_ideal-fake-real}}
\end{figure*}

The results of running the HHL circuit for the \HS flow on the IBM Nairobi device are shown in \fig\ref{fig:results_error_HS_ideal-fake-real}. The solution for the runs with and without M3 mitigation remains noisy. While this mitigation is quite effective for simulators with noise models ({\tt FakeNairobi} backend), the error mitigation strategy does not have a significant effect on real hardware for these circuits. However, these results on the real hardware are questionable since explicit modeling of IBMQ Nairobi's noise on ideal simulators did not replicate the baseline results without M3 mitigation. Two possible reasons behind this discrepancy are that the parameters in the noise model do not replicate the effects of noise at the time the HHL circuit was executed (as major calibrations of IBM devices take place only daily), and/or noise on real hardware cannot be simply described as uncorrelated depolarizing errors represented as bit-flip or phase-flip circuit instructions. Understanding the limitations of these noise models helps to bridge the gap between error-free simulator results and hardware results. More sophisticated error modeling is possible, but requires increasing computational overhead. It is very case-by-case dependent how one should trade between fidelity and cost of error modeling. In the future, we would like to explore higher fidelity noise modeling to better inform our error mitigations strategies. This would involve paying a higher classical simulation cost for improved quantum hardware results.

In pursuit of improving the accuracy of results from the hardware, we try two additional strategies to reduce the errors: (1) optimization in the transpile operation and (2) dynamic decoupling. Both these strategies are based on error suppression -- reducing the undesirable effects within the circuit by customization. As a proof of concept and brevity of the large register measurements, we demonstrate these techniques on sample linear system of equations with a matrix size of $2 \times 2$. The number of qubits required by the HHL algorithm to solve this system of linear equations is $5$.

\subsubsection{Optimization in the transpile operation -- Suppressing gate operation errors} 

One of the main sources of errors in quantum circuits is through two-qubit gate operations, and particularly through the use of swap gates, modeled using CNOT gates. Swap gates are more prevalent in superconducting quantum computers where the qubit connectivity is not all-to-all. Particularly for our HHL implementation of the \HS problem, the circuit depths are quite high with a lot of entanglement of non-neighbor qubits. 
Thus, appropriate optimization of the circuits to reduce the number of non-local gates (CNOT gates) and the circuit width is important to reduce the level of error incurred by the backends. 

Qiskit provides three levels of optimization to help reduce the number of non-local gates, total number of gates, and circuit width. Level $0$ does not perform any explicit optimization of the circuit and just transforms the circuit to match the topology of the device and its basis gates. Levels $1$, $2$, and $3$ perform light, medium, and heavy optimizations, respectively. These are done using a combination of passes which are configured to search for optimized circuits and find which gates can be collapsed. Thus, higher optimization levels generally require more time as the number of searches is higher. Level $3$ additionally performs peephole optimization by combining a chain of gates on the same qubits and re-synthesizing them with better cost.

\begin{figure}
\begin{center}  \includegraphics[width=0.5\textwidth]{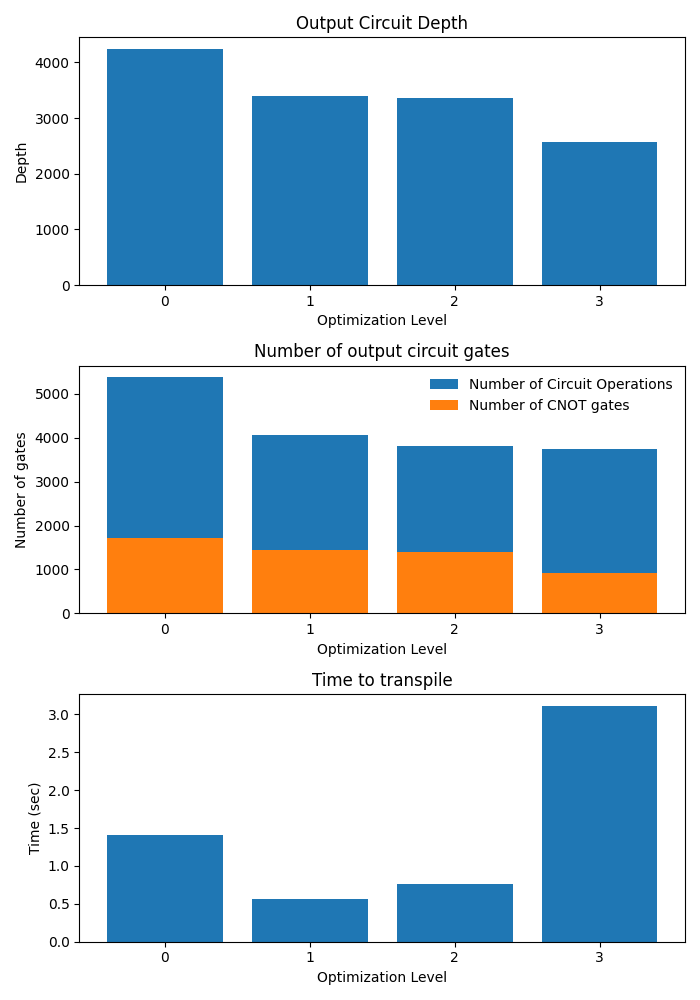}
\end{center}
 \caption{Comparing the circuit properties with change in optimization level (from $0$ to $3$) in the transpilation operation. The circuit solves a sample linear system of equations of size 2 (size of matrix $A = [2\times2]$). The circuit is transpiled for a fake backend having noise model of the IBM Cairo machine. 
\label{fig:results_transpile_opt}}
\end{figure}

The results of the different optimizations in the transpile operation are shown in \fig\ref{fig:results_transpile_opt}. Note the very large circuit depth ($>4000$) for this simple problem ($5$-qubit circuit) in the baseline case (level $0$). The various optimization levels are indeed able to reduce the circuit depth, number of CNOT gates, and the total number of gates. The time taken for transpilation at different optimization levels are interesting. While optimization level $3$ requires the longest time to complete, which is expected, optimization levels $1$ and $2$ require lesser time compared to no (or minimal) optimization (level $0$).

\subsubsection{Dynamic decoupling} 

The second strategy that we use is dynamic decoupling (DD) \cite{viola1999dynamical}. The DD technique is related to the decoherence of idle qubits when they interact with the outside environment or other nearby qubits that are used in a calculation. With increasing circuit depths, the chance of qubits idling is higher, leading to the danger of them interacting with the environment and nearby qubits, which causes decoherence. The DD technique is an error suppression technique of passing sequence of gates (amounting to identity) to idle qubits (containing delay instructions) to cancel out unnecessary interactions with the environment. Thus the procedure helps alleviate qubit decoherence during the idle time periods. We apply single-qubit Pauli-X gates to idle qubits. 

\begin{figure*}
\begin{center}
\includegraphics[width=1\textwidth]{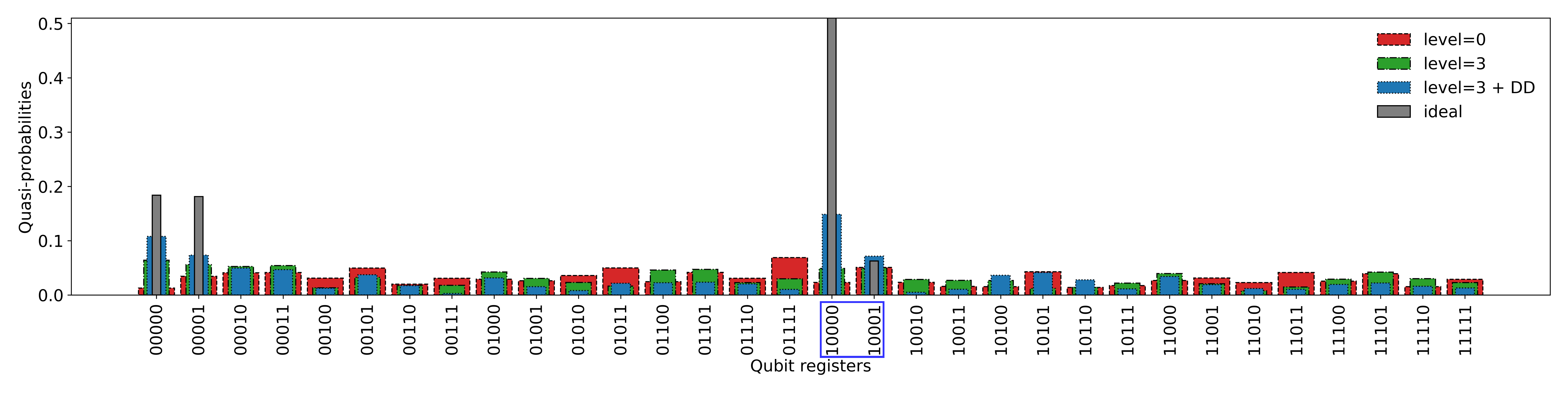}
\end{center}
 \caption{Comparing the effect of transpile optimization level and dynamic decoupling on the results of running the sample linear system of equations analyzed in \fig\ref{fig:results_transpile_opt} on the real IBM Cairo machine.
\label{fig:results_transpile_opt_dd}}
\end{figure*}

Results of running the HHL circuit for the sample problem on real hardware (IBM Cairo) with different levels of optimization and DD are shown in \fig\ref{fig:results_transpile_opt_dd}. The optimization procedures improve the overall capability of the simulations compared to no optimization, but do not significantly improve the accuracy resulting from IBM's device, even with the reduction in the number of CNOT gates. Nonetheless, incorporating the DD technique with optimization significantly increases the accuracy and reduces the noise in the measurements from IBM's device.

\section{Concluding Remarks and Outlook}
\label{sec:conclusion}

We assess the capability of a quantum linear systems algorithm (QLSA), namely the Harrow--Hassidim--Lloyd (HHL) algorithm,  to efficiently solve an idealized fluid flow problem, the Hele--Shaw problem, using superconducting devices -- IBM's Nairobi and Cairo devices. We focus on the following aspects in the quantum computing campaign: (1) accuracy of the HHL algorithm, (2) scalability of the algorithm on simulators, (3) modeling and mitigating noise in the hardware using simulators, and finally (4) results for running on superconducting devices. We use Qiskit to perform all our analyses.

The state representation of the fluid flow problem has a few bottlenecks. Usually, the Jacobian matrices resulting from the finite difference discretization of the governing equations are not Hermitian. The size of the system doubles when the matrix is transformed to be Hermitian via complex conjugate operations. Furthermore, the flexibility of solving problems with varying grid resolution is limited as the size of the system needs to be powers of $2$. Additional padding of the system matrix and vector is required to fit the problem for the QLSA algorithm, which can result in ill-conditioned systems. The Qiskit implementation of the HHL algorithm achieves very high accuracy ($>99.99\%$ fidelity) compared to previous demonstrations when solving the velocity and pressure profiles of the \HS problem on simulators. In terms of problem size, the computational cost of the implementation scales exponentially -- with circuit generation consuming the most time. Qiskit's noise models showed limitations in replicating the actual noise experienced on real hardware, leading to noisy quasi-probability measurements.

We also assess the capability of various error mitigation and suppression strategies to alleviate the erroneous results from the device. The error mitigation strategy is focused on mitigating qubit measurement errors, tested using the matrix-free measurement mitigation (M3) technique implemented in Qiskit. While the M3 technique helps improve the accuracy of the simulators with noise models, the results from real hardware do not show improvement. We also test error suppression techniques to alleviate error from gate operations and qubit decoherence. Various optimization techniques in the transpilation operation are used to reduce the number of swap gates (CNOT gates), total number of gates, and the circuit depth. Dynamic decoupling (DD) is applied to mitigate decoherence of idling qubits due to interaction with the outside environment and neighboring qubits in the very deep HHL circuits generated for the superconducting devices. We find that incorporating DD with high circuit optimization significantly increases the accuracy when solving small system sizes.

The inconsistency of Qiskit's noise models compared to real hardware is expected, as these models are generated using snapshots of the device in the past and are not representative of the current state of the device. Thus, future work should focus on more representative noise models of the devices. While DD effectively alleviates errors for a small sample system, its effectiveness needs to be evaluated on larger systems and circuit widths representative of fluid flow problems. Finally, the results from our study reveal that qubit connectivity in the devices is a crucial factor in effectively solving systems of linear equations. Therefore, future studies should explore trapped-ion devices with all-to-all connectivity -- which eliminates the need for swap gates. The preliminary assessments in this study are important for more sophisticated studies aimed at evaluating the quantum utility \cite{herrmann2023quantum} of QLSA in solving fluid flow problems.


\section*{Acknowledgements}
The authors thank the insightful discussions with Pooja Rao, Dmitry Lyakh, In-Saeng Suh, and Matt Norman. This research used resources of the Oak Ridge Leadership Computing Facility at the Oak Ridge National Laboratory, which is supported by the Office of Science of the U.S. Department of Energy under Contract No. DE-AC05-00OR22725.

\section*{Data and code availability}
The specific data and codes are openly accessible at \cite{gopalakrishnan_meena_2024_13738192}.

\section*{Author contributions}
M.G.M, K.G, J.L., and A.G., initiated and designed the project. M.G.M led the formal analysis with contributions from all authors. K.G. provided help with analyzing the flow problem and setting up the linear solvers. J.L. and A.G. provided help with the computational cost studies. J.L and E.A.C.P provided guidance and help with the quantum computing assessments.

\section*{Author declaration}
The authors declare no competing interests.


\bibliographystyle{unsrtnat}
\bibliography{references}

\end{document}